\documentclass[12pt]{article}
\usepackage{epsf}
\def\la{\mathrel{\mathpalette\fun <}}

\def\fun#1#2{\lower3.6pt\vbox{\baselineskip0pt\lineskip.9pt
\ialign{$\mathsurround=0pt#1\hfil
##\hfil$\crcr#2\crcr\sim\crcr}}}
\newcommand{\vep}{\mbox{\boldmath${\rm p}$}}

\newcommand{\veR}{\mbox{\boldmath${\rm R}$}}
\newcommand{\veL}{\mbox{\boldmath${\rm L}$}}
\newcommand{\veS}{\mbox{\boldmath${\rm S}$}}
\newcommand{\veB}{\mbox{\boldmath${\rm B}$}}

\newcommand{\ver}{\mbox{\boldmath${\rm r}$}}

\newcommand{\vePsi}{\mbox{\boldmath${\rm \Psi}$}}
\newcommand{\be}{\begin{equation}}
\newcommand{\ee}{\end{equation}}
\newcommand{\bc}{\begin{center}}
\newcommand{\ec}{\end{center}}
\newcommand{\lt}{\left}
\newcommand{\rt}{\right}

\newcommand{\lb}{\label}

\newcommand{\mr}[1]{\mathrm{#1}}

\newcommand{\fr}[2]{\frac{#1}{#2}}

\newcommand{\LS}{\mathrm{LS}}

\title{\bf Adiabatic potentials and spectra of heavy hybrid mesons}
\author{Yu.S.Kalashnikova and D.S.Kuzmenko}
\date{\it Institute of Theoretical and Experimental
Physics,\\ 117218, B.Cheremushkinskaya 25, Moscow, Russia}

\begin{document}
\maketitle

\begin{abstract}
Using the QCD string approach the adiabatic potentials and spectra 
of $b\bar b$-hybrid mesons are calculated. The results are 
compared to lattice studies.
\end{abstract} 

1. An assumption about existence of the hybrid mesons consisting
of valence quark and antiquark and valence gluon was made
for the first time by Okun and Vainshtein \cite{Okun} just 
afterwards the QCD appeared. Last decade was a time of intensive  
studies of hybrid mesons both in experimental and in theoretical
frameworks.
 
In this report we study the system of static quark and antiquark
and dynamical gluon joined to the formers by the infinitely thin
string of background gluon field in analytic QCD string approach
\cite{Simonov}, \cite{3}  based on the background perturbation 
theory \cite{1}. In other words, we study spectrum of the 
vibrations of the confining string, or spectrum of the adiabatic 
potentials. We will pay our attention to the small and intermediate 
quark-antiquark separations, which are directly related to the 
spectrum of heavy hybrid mesons. Using the adiabatic potentials, we 
will calculate spectra of $b\bar b$-hybrid mesons.

2. In the framework of the background perturbation theory we
start from the propagator of the valence gluon in the background 
field $B$ \cite{1},
\be
G_{\mu\nu} = (D^2(B)\,\delta_{\mu\nu}+ 2igF_{\mu\nu}(B))^{-1},
\lb{1}
\ee
where $D(B)$ is the covariant derivative depending on the field 
$B$; $F_{\mu\nu}(B)$ is the background field strength tensor, 
which is related to the spin effects and will be considered below. 
In the formalism of the Fock-Feynman-Shwinger representation 
 the Green function of the hybrid meson with the static 
quark and antiquark reduces to the form (see \cite{3}  and 
references therein) 
\be
 G_h(\bar X, X)= \int^\infty_0 ds \int({\cal D}z_g)_{x_gy_g} \exp 
(-K) \langle {\cal W}_h\rangle_B,
\lb{2}
\ee
where ${\cal W}_h$ is the Wilson loop of the hybrid meson,
consisting of the trajectories $\Gamma_Q$, $\Gamma_{\bar Q}$
and  $\Gamma_g$ of valence quark, antiquark and gluon (see Fig. 1), 
and $K$ is the kinetic energy of the gluon.

\begin{figure}[!t]
\epsfxsize=8cm
\hspace*{4.35cm}
\epsfbox{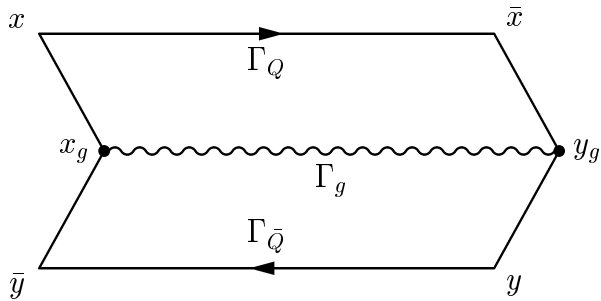}
\caption{Wilson loop of hybrid meson}
\end{figure}

In what follows we will use the area law
for the Wilson loop, which is confirmed by the lattice QCD, see 
e.g. \cite{2}. The area law for hybrid mesons takes the form 
\be
\langle {\cal W}_h\rangle_B =
\frac{N_c^2-1}{2}\exp (-\sigma(S_1+S_2)),
\lb{3}
\ee
where $S_1~(S_2)$ is an area of the minimal surface bounded by the 
trajectores of valence quark (antiquark) and gluon. 
We parametrise the minimal surfaces introducing
parameters $0\leq \tau\leq T$ and $0\leq \beta_{1,2}\leq 1$.
It is also convenient to introduce einbein fields $\mu(\tau),~
\nu_{1,2}(\tau,\beta)$ and write the Green function of the hybrid 
meson in the form 
\be
G_h(\bar X, X)= \int({\cal D}^3 z_g)_{x_gy_g}{\cal D}\mu\,{\cal
D}\nu_1\, {\cal D}\nu_2\exp \left(-\int_0^T 
d\tau\,L\right),
\lb{4}
\ee
which defines the following string Hamiltonian \cite{3},
\be
H={\vep}\, \dot{\ver} - L=H_0+\frac{\mu}{2}
+\int^1_0 d\beta_1\,\frac{\sigma^2 r^2_1}{2\nu_1}
+\int^1_0
d\beta_2\,\frac{\sigma^2 r^2_2}{2\nu_2}+\int^1_0
d\beta_1\,\frac{\nu_1}{2}+
\int^1_0
d\beta_2\,\frac{\nu_2}{2},
\lb{5}
\ee
where
$$
H_0=\frac{p^2}{2(\mu+J_1+J_2)}+
\frac{1}{2\Delta (\mu+J_1+J_2)}\times
$$
\be
\times
\left\{ \frac{(\vep\,\ver_1)^2}{r^2_1}J_1(\mu+J_1)+
\frac{(\vep\,\ver_2)^2}{r^2_2}J_2(\mu+J_2)+
\frac{2J_1J_2}{r^2_1r^2_2}(\ver_1\ver_2)(\vep\,\ver_1)(\vep\,\ver_2)
 \right\},
\lb{6}
\ee
\bigskip
$$
\Delta=(\mu +J_1) (\mu+J_2) -
J_1J_2\frac{(\ver_1\ver_2)^2}{r^2_1r^2_2},~~
J_i=\int^1_0 d\beta_i\,\beta_i^2\nu_i(\beta_i),~~ i=1,2,
$$
$\ver$ and $\vep$ are the coordinate and momentum of the valence 
gluon, $\ver_{1,2}=\ver\pm \veR/2$, where $\pm\veR/2$ are
locations of the static quark and antiquark. Einbein fields
are to be excluded from the Hamiltonian by the conditions
\be
\frac{\partial H}{\partial\mu}=0,~~
\frac{\delta H}{\delta \nu_i(\beta_i)}=0.
\lb{7}
\ee

One can show that einbein $\mu$ plays the 
role of the constituent mass of the gluon, $\nu_i(\beta)$ the 
energy density of the background field along the string, and $J_i$ 
the inertia of the string.
 
3. Proceed now to the calculations of the spectrum of the string 
Hamiltonian at small and intermediate quark-antiquark separations,
$R\la 1$ fm. In this region  the inertia of the 
string $J_i$ is much smaller than the constituent mass of the 
gluon, $J_i\ll \mu$. Neglecting $J_i$ and calculating extremum 
over $\mu$ in the initial Hamiltonian, we  arrive at the 
Hamiltonian of the potential model, 
\be
H=\sqrt{\vep^2}+\sigma r_1+\sigma r_2.
\lb{8}
\ee
However, it is more convenient to calculate the eigenvalues 
$E(\mu)$ of the Hamiltonian keeping the einbein $\mu$, 
and then minimize them with respect to the einbein,
\be
\frac{\partial E(\mu)}{\partial \mu}=0.
\lb{9}
\ee
 This procedure is of common use in the QCD string 
approach,  and, for the ordinary mesons, is justified with the 
accuracy better than 5\% (see e.g. \cite{Simonov}). 

Besides the nonperturbative string part considered above,
the hybrid Hamiltonian should contain the perturbative one $V_c$,
which is dominated by the one-gluon exchange,
\be
V_c=-\fr{3\,\alpha_s}{2r_1}-\fr{3\,\alpha_s}{2r_2}+\fr{\alpha_s}{
6R}.  
\lb{10}
\ee
Therefore, the  Hamiltonian with the string inertia neglected
takes the form
\be
H=\frac{\mu}{2}+\frac{p^2}{2\mu}+\sigma r_1+\sigma r_2+V_c.
\lb{11}
\ee
Despite its nonrelativistic appearance, this Hamiltonian is 
essentially relativistic due to condition (\ref{9}). 

The Hamiltonian of the string correction of the order of $J/\mu$
derived from (\ref{5}), (\ref{6}) has transparent form,
\be
H^{\mathrm{string}}=-\frac{\sigma}{6\mu^2}\left(
\frac{1}{r_1}L_1^2+\frac{1}{r_2}L_2^2\right),
\lb{12}
\ee
 where $\veL_i=\ver_i\times\vep$.

Spin interaction in ordinary mesons with heavy quarks of mass 
$m$ have been expressed through the number of universal uknown 
potentials of order $1/m^2$ using the Wilson loop in \cite{Gromes}. 
In the framework of QCD string approach these potentials were 
calculated using the method of field 
correlators, see \cite{Simonov} and references therein. 
The role of the mass $m$ is performed in the approach by 
the einbein   $\mu$, and the results therefore are applicable to 
both light and heavy quarks.
 
Spin interaction of the valence gluon with the quark and 
antiquark in hybrid mesons is generated by the strength tensor 
term in propagator (\ref{1}), which can be rewritten as
$F_{ik}=i(\veS \veB)_{ik}$, where the spin operator $\veS$
acts onto the gluon wave function $\vePsi$ according to 
$\left(S_i \vePsi\right)_j = -i\epsilon_{ijk}\Psi_k$.
Performing calculations similar to ones of \cite{Simonov},
we derive the following relations in Eichten-Feinberg
notations,
\be
V_1^{\mr{(p)}}=0,~~V_2^{\mr{(p)}}=\epsilon^{\mr{(p)}},\quad
\mbox{ where }\quad \epsilon^{\mr{(p)}}=-\fr{3\alpha_s}{2r_i}, 
~~i=1,2, 
\lb{GR1}
\ee
and
\be
V_1^{\mr{(np)}}=-\epsilon^{\mr{(np)}},~~V_2^{\mr{(np)}}=0,\quad
\mbox{ where }\quad \epsilon^{\mr{(np)}}=\sigma r_i.
\lb{GR2}
\ee
One can verify that the Gromes relation $\epsilon +V_1-V_2=0$
is justified  for both perturbative and nonperturbative 
potentials. The corresponding Hamiltonian consists  of the 
following perturbative and nonperturbative parts \cite{3}, 
\be
H^{\LS\,\mr{(p)}}= \frac{3\alpha_s}{4\mu^2} \left(
\frac{\veL_1 \veS}{r_1^3} +\frac{\veL_2 \veS}{r_2^3}
\right),
\lb{13}
\ee
\be
H^{\LS\,\mr{(np)}}= -\frac{\sigma}{2\mu^2} \left(
\frac{\veL_1 \veS}{r_1} +\frac{\veL_2 \veS}{r_2}
\right).
\lb{14}
\ee

We calculate the energy levels of the initial Hamiltonian
(\ref{11}) variationally, using the Gaussian wave functions
of the gluon,
\be
\vec\Psi_{jl\Lambda}(\ver)=\phi_l(r)\sum_{\mu_1\mu_2}
C^{j\Lambda}_{l\mu_1 l\mu_2}Y_{l\mu_1}\left(\frac{\ver}{r}\right)
\vec\chi_{1\mu_2},
\lb{15}
\ee
$$
\phi_l(r)\sim \exp\lt(-\beta^2 r^2/2\rt).
$$
The corrections (\ref{12})-(\ref{14}) amount to the decrease of 
the total energy about 70-150 MeV depending on the level.

\begin{figure}[!t]
\epsfxsize=15cm
\centering
\vspace*{0.85cm}
\epsfbox{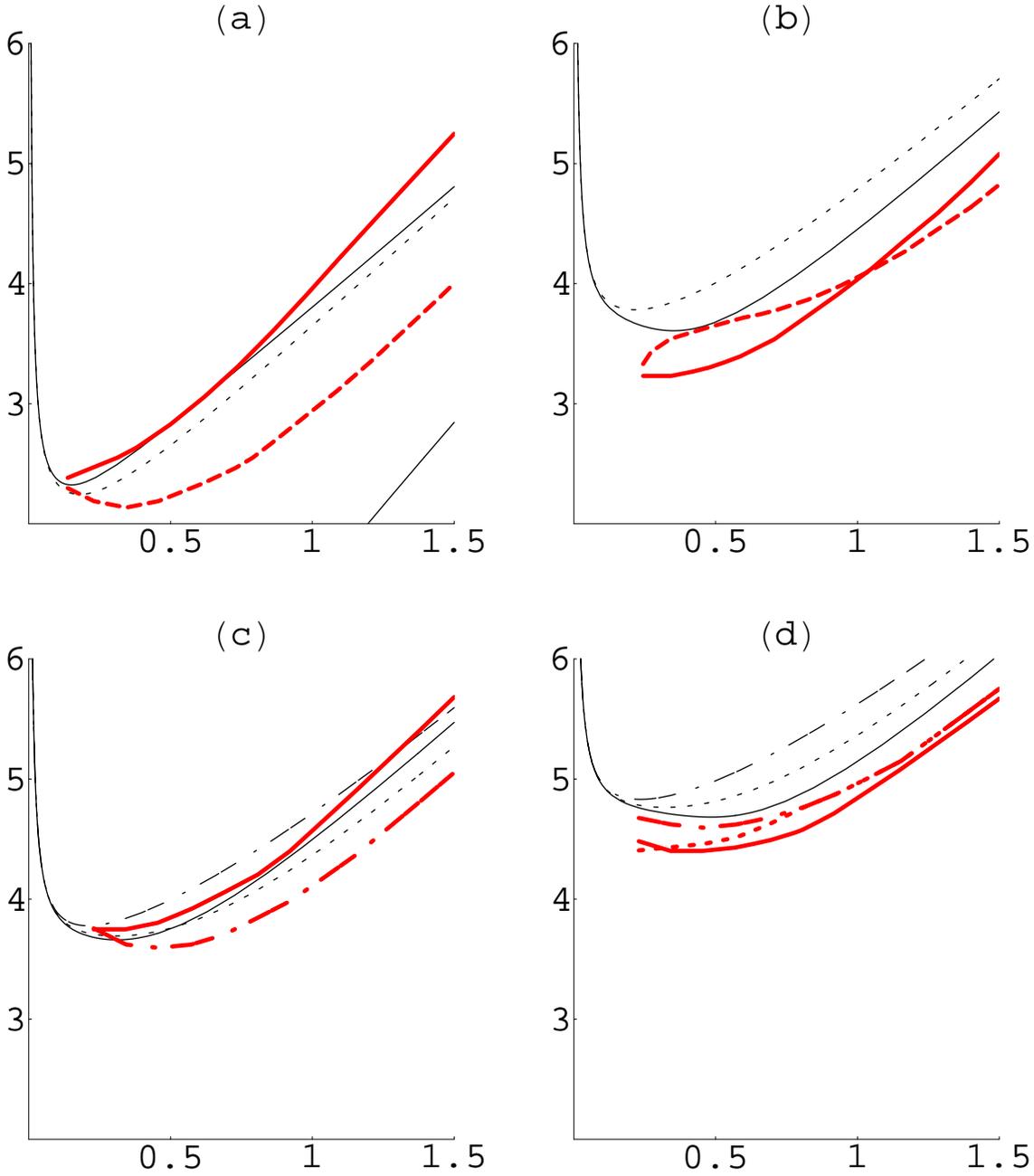}
\caption{Hybrid potentials with corrections included (black curves)
compared to lattice ones (color curves)
in units $1/r_0=$ 400 MeV ($r_0=2.5$ GeV$^{-1}$).  $Q \bar
Q$-distance $R$ is measured in $2r_0\approx$ 1 fm. States
(a)--(d) are given in Table \ref{qnum}. Solid, dashed and
dashed-dotted curves correspond to $\Lambda=0,1$ and 2. Solid line
at right bottom corner of Fig.\ref{pot}(a) represents the
Coulomb+linear potential with $\alpha_s=0.4$ and $\sigma=0.21$
GeV$^2$.}
\lb{pot}
\vspace*{2cm}
\end{figure}

\begin{table}[!t]
\centering
\label{table}
\begin{tabular}{|c l|}
  \hline \rule{0cm}{.7cm}
  (a) &   $\Sigma_u^-,~\Pi_u$ \\[5pt]
  (b) &   $\Sigma_g^+,~\Pi_g$ \\[5pt]
  (c) &   $\Sigma_g^-,~\Pi_g,~\Delta_g$ \\[5pt]
  (d) &   $\Sigma_u^+,~\Pi_u,~\Delta_u$ \\[5pt]
\hline
  \end{tabular}
\caption{Quantum numbers of adiabatic levels of Fig. 2}
\label{qnum}
\end{table}

In Fig. 2 the resulting adiabatic potential energy levels   
are shown in comparison with the lattice results \cite{4}.   
Quantum numbers of the levels in the  diatomic 
molecule notations are given in Table 1. The potentials are normalized to 
the ground $Q\bar Q$-level shown in the right bottom corner of 
Fig. 2(a). We can see overall correspondence with the lattice. The 
curves  tend to form three multiplets with different values of the 
gluon angular momentum. This is a confirmation that the angular 
moment of the gluon is a good quantum number. When $R$ goes to 
zero, the potentials increase rapidly due to the Coulomb repulsion 
of  quark and antiquark in octet state of color $SU(3)$. Note the 
wrong bend of $\Pi_g$ lattice level in Fig. 2(b) and absence of  
$\Pi_g$ lattice level in Fig. 2(c), which  clearly means that 
these two  $\Pi_g$  levels are not properly resolved on the 
lattice. The order of levels in our approach correspond to the 
lattice one, with the exception of Fig. 2(c). Different  
separations between the levels of Fig. 2(a) in our and lattice 
approaches is presumably related to the influence of the Coulomb 
gluon-quark and gluon-antiquark interaction on the gluon wave 
function. Variational calculations using the
Coulomb-modified wave functions are needed to verify 
this assumption, which will be performed in future publications.
 
Lowest level with the quantum numbers $\Sigma^-_g,~\Pi_g$ is  
absent on the lattice and therefore not shown in Fig. 2.
Comparative analysis of adiabatic 
potentials of other approaches can be found in \cite{3}.

4. Let us calculate spectra of masses of the heavy hybrid mesons
in Born-Oppenheimer approximation assuming that the motion of 
the valence quark and antiquark is slow compared to the motion of 
the valence gluon and using the adiabatic potentials.
In this case an essential region is the one near the minimum of the 
potential, where the adiabatic potential can be replaced by the 
oscillatory one. Therefore the problem reduces to the calculation
of the spectrum of three-dimentional oscillator. 

We tune the  values of parameters $\alpha_s=0.225$, 
$\sigma=0.18$ GeV$^2$, $m_b=4.56$ GeV to reproduce
the experimental value of the ground 
state of  $b \bar b$-meson,  $M^{\Upsilon(1S)}_{\mr{exp}}=9.4$ 
GeV, from the Cornell potential, 
\be
V^{b\bar b}=2m_b-\frac43 \frac{\alpha_s}{R}+\sigma R.
\lb{16}
\ee
 Then we calculate 
the masses of $b \bar b$ hybrid mesons through the equation
\be
M^{b\bar b g}=2m_b+E^{\mr{osc}},
\lb{17}
\ee
where $E^{\mr{osc}}$ is the energy of $b$-quarks in the 
adiabatic potential, 
approximated in the vicinity of its minimum by the oscillatory 
one. 
\begin{table}[!t]
\centering
\begin{tabular}{|l | l|}
  \hline \rule{0cm}{.7cm}
  $l=1,~l_{Q\bar Q}=0$ (GE) &
   $M^{b\bar b g}=10.83 \mbox{ 
GeV}$ \\[5pt]
  $l=0,~l_{Q\bar Q}=1$ (QE) &
    $M^{b\bar b g}=10.92\mbox{ GeV}$\\[5pt]
\hline
  \end{tabular}
\caption{The masses of the lowest $b\bar b$-exotic states in  our 
approach} 
\label{masses}
\end{table}

\begin{table}[!t]
\centering
\begin{tabular}{|l | l|}
  \hline \rule{0cm}{.7cm}
Lattice QCD \cite{4}&  $M^{b\bar b g}$=10.8 GeV \\[5pt]
QCD string approach \cite{5}&  $M^{b\bar b g}$=10.64 GeV
\\[5pt]
\hline
  \end{tabular}
\caption{The masses of lowest $b\bar b$-exotic states with 
$J^{PC}=1^{-+}$ in other works} 
\label{table3}
\end{table}

The lowest exotic  hybrid states have quantum numbers 
$J^{PC}=1^{-+}$ and may be of two different kinds,  the 
gluon-excited (GE) one, with the gluon angular momentum $l=1$ 
and quark-antiquark relative angular moment $l_{Q\bar Q}=0$, and 
quark-excited (QE) one, with $l=0$ and $l_{Q\bar Q}=1$. 
Their masses are given in Table 2.  One can calculate the energy 
of the gluon excitation using the mass of the GE-state from the 
Table,
\be 
\Delta M=M^{b\bar b g}_{(GE)}-M^{\Upsilon(1S)}=1.43 \mbox{ GeV}. 
\lb{18}
\ee
One can see that $\Delta M\ll m_b$ so that the adiabatic 
approximation for $b\bar b$-hybrid  mesons is justified.
The mass of $c$-quark is close to the energy of 
gluon excitation, so that adiabatic approximation
for $c\bar c$-hybrid mesons is invalid.

 The mass of the lowest exotic state
of $b \bar b$-hybrid meson was calculated  on the lattice 
\cite{4}, as well as in the QCD string approach  in the  
work \cite{5}. The results  are shown  in the Table 3. 
In lattice QCD lowest quark-excited state is absent,
as it was already noted above. One can see from the Tables 2, 
3 that the mass of the gluon-excited state computed in the 
adiabatic approximation is almost the same as  ours. Note that  in 
both cases quenched approximation was used. One can guess 
that a proper account of the see quarks will not change the result 
significantly. 

The results of the work \cite{5} were 
obtained using the Hamiltonian (\ref{11}) and the method of 
einbein field. 
The main difference between this and ours results is due to
the large constant  subtracted in  \cite{5}. 
The adiabatic Born-Oppenheimer approximation was not used in 
 \cite{5} and string- and spin-corrections to Hamiltonian were 
not considered. 

One can find recent comparison of existing calculations
of lowest exotic hybrid meson masses in other models in 
\cite{iddir}. Note  that ITEP sum rules and flux-tube model 
predict lowest exotic $b\bar b$ hybrid mass with large error bars, 
which include both our numbers, and bag model seems to 
underestimate the mass. Recent calculations of GE-mode
in quark-gluon model \cite{iddir} are in good agreement with both 
lattice and our results. The mass of QE-mode in  \cite{iddir}
is less than GE-one, unlike ours, see Table 2. This discrepency is
presumably due to poor accuracy of our variational procedure
for lowest adiabatic level.

In conclusion, we have shown that the QCD string approach allows 
to calculate the adiabatic potentials and masses of heavy hybrid 
mesons in a good agreement with lattice QCD. Our further
tasks are use of more elaborated variational procedure for the 
calculation of the lowest adiabatic levels, as well as accurate 
calculations of hybrid spectra, wave functions and decays
both within and beyond the adiabatic approximation. 

This work has been supported by 
 RFBR grants 00-02-17836,  00-15-96786, and INTAS 00-00110.

\end{document}